\documentstyle[12pt,epsf]{article}
\newlength{\pubnumber} 

\catcode`\@=11
\@addtoreset{equation}{section}

\def\section{\@startsection{section}{1}{\z@}{3.5ex plus 1ex minus .2ex}
 {2.3ex plus .2ex}{\large\bf}}
\def\subsection{\@startsection{subsection}{2}{\z@}{2.3ex plus .2ex}
 {2.3ex plus .2ex}{\bf}}

\input epsf

\begin{document}

\begin{titlepage}
\samepage{
\setcounter{page}{1}
\rightline{OUTP--04--01P}
\rightline{\tt hep-th/0403272}
\rightline{January 2004}
\vfill
\begin{center}
 {\Large \bf On the Number of \\Chiral Generations in
$Z_2\times Z_2$ Orbifolds }
\vfill
\vfill {\large Ron Donagi$^{1}$\footnote{donagi@math.upenn.edu} 
and 
               Alon E. Faraggi$^{2}$\footnote{faraggi@thphys.ox.ac.uk}}\\
\vspace{.12in}
{\it $^{1}$ Department of Mathematics,
            University of Pennsylvania,\\
            Philadelphia, PA, 19104--6395, USA\\}
\vspace{.05in}
{\it $^{2}$ Theoretical Physics Department, University of Oxford,
            Oxford, OX1 3NP, UK\\
	    and \\
	    School of Natural Sciences, Institute for Advanced Study,\\
Princeton, NJ 08540, USA\\}
\vspace{.025in}
\end{center}
\vfill
\begin{abstract}

The data from collider experiments and cosmic observatories
indicates the existence of three light matter generations.
In some classes of string compactifications the 
number of generations is related to a topological
quantity, the Euler characteristic.
However, these do not explain the existence of three
generations. In a class of free fermionic string models,
related to the $Z_2\times Z_2$ orbifold compactification,
the existence of three generations is correlated
with the existence of three twisted sectors
in this class of compactifications.
However, the three generation models are constructed
in the free fermionic formulation and their
geometrical correspondence is not readily available.
In this paper we classify quotients of the $Z_2\times Z_2$ 
orbifold by additional symmetric shifts on
the three complex tori. We show that three
generation vacua are not obtained in this manner,
indicating that the geometrical structures
underlying the free fermionic models are more
esoteric.

\end{abstract}
\smallskip}
\end{titlepage}

\setcounter{footnote}{0}

\def\beq{\begin{equation}}
\def\eeq{\end{equation}}
\def\beqn{\begin{eqnarray}}
\def\eeqn{\end{eqnarray}}

\def\no{\noindent }
\def\nolabel{\nonumber }
\def\ie{{\it i.e.}}
\def\eg{{\it e.g.}}
\def\half{{\textstyle{1\over 2}}}
\def\third{{\textstyle {1\over3}}}
\def\quarter{{\textstyle {1\over4}}}
\def\sixth{{\textstyle {1\over6}}}
\def\m{{\tt -}}
\def\p{{\tt +}}

\def\Tr{{\rm Tr}\, }
\def\tr{{\rm tr}\, }

\def\slash#1{#1\hskip-6pt/\hskip6pt}
\def\slk{\slash{k}}
\def\GeV{\,{\rm GeV}}
\def\TeV{\,{\rm TeV}}
\def\y{\,{\rm y}}
\def\SM{Standard--Model }
\def\SUSY{supersymmetry }
\def\SSSM{supersymmetric standard model}
\def\vev#1{\left\langle #1\right\rangle}
\def\l{\langle}
\def\r{\rangle}
\def\o#1{\frac{1}{#1}}

\def\Htw{{\tilde H}}
\def\chibar{{\overline{\chi}}}
\def\qbar{{\overline{q}}}
\def\ibar{{\overline{\imath}}}
\def\jbar{{\overline{\jmath}}}
\def\Hbar{{\overline{H}}}
\def\Qbar{{\overline{Q}}}
\def\abar{{\overline{a}}}
\def\alphabar{{\overline{\alpha}}}
\def\betabar{{\overline{\beta}}}
\def\tautwo{{ \tau_2 }}
\def\thetatwo{{ \vartheta_2 }}
\def\thetathree{{ \vartheta_3 }}
\def\thetafour{{ \vartheta_4 }}
\def\ttwo{{\vartheta_2}}
\def\tthree{{\vartheta_3}}
\def\tfour{{\vartheta_4}}
\def\ti{{\vartheta_i}}
\def\tj{{\vartheta_j}}
\def\tk{{\vartheta_k}}
\def\calF{{\cal F}}
\def\smallmatrix#1#2#3#4{{ {{#1}~{#2}\choose{#3}~{#4}} }}
\def\ab{{\alpha\beta}}
\def\Minv{{ (M^{-1}_\ab)_{ij} }}
\def\bone{{\bf 1}}
\def\ii{{(i)}}
\def\V{{\bf V}}
\def\N{{\bf N}}

\def\b{{\bf b}}
\def\S{{\bf S}}
\def\X{{\bf X}}
\def\I{{\bf I}}
\def\mb{{\mathbf b}}
\def\mS{{\mathbf S}}
\def\mX{{\mathbf X}}
\def\mI{{\mathbf I}}
\def\balpha{{\mathbf \alpha}}
\def\bbeta{{\mathbf \beta}}
\def\bgamma{{\mathbf \gamma}}
\def\bxi{{\mathbf \xi}}

\def\t#1#2{{ \Theta\left\lbrack \matrix{ {#1}\cr {#2}\cr }\right\rbrack }}
\def\C#1#2{{ C\left\lbrack \matrix{ {#1}\cr {#2}\cr }\right\rbrack }}
\def\tp#1#2{{ \Theta'\left\lbrack \matrix{ {#1}\cr {#2}\cr }\right\rbrack }}
\def\tpp#1#2{{ \Theta''\left\lbrack \matrix{ {#1}\cr {#2}\cr }\right\rbrack }}
\def\l{\langle}
\def\r{\rangle}


\def\inbar{\,\vrule height1.5ex width.4pt depth0pt}

\def\IC{\relax\hbox{$\inbar\kern-.3em{\rm C}$}}
\def\IQ{\relax\hbox{$\inbar\kern-.3em{\rm Q}$}}
\def\IR{\relax{\rm I\kern-.18em R}}
 \font\cmss=cmss10 \font\cmsss=cmss10 at 7pt
\def\IZ{\relax\ifmmode\mathchoice
 {\hbox{\cmss Z\kern-.4em Z}}{\hbox{\cmss Z\kern-.4em Z}}
 {\lower.9pt\hbox{\cmsss Z\kern-.4em Z}}
 {\lower1.2pt\hbox{\cmsss Z\kern-.4em Z}}\else{\cmss Z\kern-.4em Z}\fi}

\def\AEF{A.E. Faraggi}
\def\NPB#1#2#3{{\it Nucl.\ Phys.}\/ {\bf B#1} (#2) #3}
\def\PLB#1#2#3{{\it Phys.\ Lett.}\/ {\bf B#1} (#2) #3}
\def\PRD#1#2#3{{\it Phys.\ Rev.}\/ {\bf D#1} (#2) #3}
\def\PRL#1#2#3{{\it Phys.\ Rev.\ Lett.}\/ {\bf #1} (#2) #3}
\def\PRT#1#2#3{{\it Phys.\ Rep.}\/ {\bf#1} (#2) #3}
\def\MODA#1#2#3{{\it Mod.\ Phys.\ Lett.}\/ {\bf A#1} (#2) #3}
\def\IJMP#1#2#3{{\it Int.\ J.\ Mod.\ Phys.}\/ {\bf A#1} (#2) #3}
\def\nuvc#1#2#3{{\it Nuovo Cimento}\/ {\bf #1A} (#2) #3}
\def\RPP#1#2#3{{\it Rept.\ Prog.\ Phys.}\/ {\bf #1} (#2) #3}
\def\etal{{\it et al\/}}

\hyphenation{su-per-sym-met-ric non-su-per-sym-met-ric}
\hyphenation{space-time-super-sym-met-ric}
\hyphenation{mod-u-lar mod-u-lar--in-var-i-ant}


\setcounter{footnote}{0}
\section{Introduction}

One of the important clues in the quest for the unification  of the 
elementary matter and interactions is the triple replication of the
Standard Model fermion states. While the possibility exists that there
are additional families, contemporary data suggests the existence
of only three chiral generations. The precision electroweak data
obtained at LEP and SLC show that the width of the $Z$--boson
can only accommodate three light left--handed neutrinos \cite{lep}. The 
constraints from observations of light element abundances 
also constrain the number of relativistic degrees of freedom
during primordial nucleosynthesis to correspond to three light
left--handed neutrinos \cite{bigbang}.
Similarly, existence of three quark generations is 
compatible with the constraints arising from unitarity of the
Cabbibo--Kobayashi--Maskawa mixing matrix. In the context 
of grand unification, gauge coupling unification and the mass
ratio $m_b/m_\tau$ are only compatible with the low energy data
in the presence of three chiral generations \cite{threeguts}.
Understanding the origin of the number of flavors and of their mass
and mixing spectrum is therefore one of the vital issues
in the phenomenology of the Standard Model and unification.

In the context of point quantum field theories the number of
generations and the flavor variables are mere parameters that
fit the data. It is plausible that understanding of the origin
of these fundamental constants can only be obtained in the 
framework of quantum gravity, {\it i.e.} their origin
is of a geometrical characteristic. It is then encouraging that
in the context of heterotic string theories \cite{heterotic}
compactified on Calabi--Yau manifolds \cite{canwit}
the number of generation in the low
energy spectrum is dictated by a topological quantity, the
Euler number $\chi$.
However, the Euler number of a random Calabi--Yau manifold
can take many values and therefore does not yet
provide a compelling explanation for the existence of three
chiral generations.

In this context it has been suggested that a particular class
of Calabi--Yau manifolds may provide a plausible insight to 
the existence of three chiral generations \cite{nahe}. The relevant 
manifolds are those related to the $Z_2\times Z_2$ orbifolds \cite{foc}
of six dimensional toroidal spaces \cite{Narain,dhvw},
that have been studied
most extensively in the free fermionic formulation of the 
heterotic string in four dimensions \cite{fff,rffm}.
The origin of the 
number of three generations in the $Z_2\times Z_2$ orbifold
is associated with the existence of three twisted sectors and
the fact that each of the three $Z_2$ twists
leaves one torus fixed. The enumeration of the number
of generations then corresponds to the number of fixed
points on the two twisted tori. The realization of the
three generations in the free fermionic models then 
corresponds to reducing the number of generations
to one generation from each of the twisted sectors
of the $Z_2\times Z_2$ orbifold. Thus, this class
of string compactifications correlates the existence of
three generations with the structure of the underlying
$Z_2\times Z_2$ compactified manifold \cite{nahe,foc}. 

The free fermionic models, however, are a particular realization 
of the $Z_2\times Z_2$ orbifold at a fixed point in the
moduli space. Furthermore, the geometrical correspondence
is well established only in specific cases and is lacking
in the case of the three generation models.
For many of the issues pertaining to the phenomenology of these
models, it will be beneficial to abandon the fermionic 
realization and to resort to the bosonic, or geometric
description. It is therefore particularly important
to understand the precise geometrical structure of the
three generation models.

The aim of this paper is therefore to study the question of the 
number of generations in $Z_2\times Z_2$ orbifolds with symmetric
shifts. The goal is to examine whether such constructions 
can reproduce the free fermionic picture of obtaining one 
chiral generation from each twisted sector. While string theory,
in general, and its free fermionic formulation, in particular, 
allows more general operations, {\it i.e.} those that are asymmetric
between the left-- and right--moving coordinates on the world--sheet,
the restriction to symmetric shifts may be viewed as what is allowed
by ``classical geometry''. In this respect our conclusion will be in
fact negative. Namely, we will prove that it is not possible to
produce the three generation manifolds solely by utilizing symmetric
shifts. This, in our view, is a substantial outcome with important
possible consequences. First, it indicates that the geometry underlying
the three generation free fermionic models is not ``classical geometry''
as it necessarily involves operations that are not symmetric
between the left-- and the right--moving coordinates. The
relevant geometrical structures may therefore be of intrinsic ``quantum''
or ``stringy'' character. Second, the fact that the three generation models
necessarily employ asymmetric operations may prove to 
be important for the question of moduli stabilization.

Our paper is organized as follows. In section \ref{gs}
we discuss how three generations are obtained in the free fermionic 
models, which serves as our motivation for the ensuing analysis.
In section \ref{z2z2corres} we discuss the free fermion--orbifold
correspondence and set the ground for the subsequent analysis.
In sections \ref{lowrankorbifolds}--\ref{beyondz2z2}
we present the complete analysis of
the $Z_2\times Z_2$ with symmetric shifts. We identify the
geometric condition for producing chiral matter,
and present the proof that $Z_2\times Z_2$ orbifold with
solely symmetric shifts cannot yield a three generation
vacuum. Section \ref{disco} concludes the paper.

\setcounter{footnote}{0}
\section{Three generations in the free fermion models}\label{gs}
In the free fermionic formulation of the heterotic string
in four dimensions all the world--sheet
degrees of freedom  required to cancel
the conformal anomaly are represented in terms of free fermions.
For the left--movers one has the
usual space--time fields $X^\mu$, $\psi^\mu$, ($\mu=0,1,2,3$),
and in addition the following eighteen real free fermion fields:
$\chi^I,y^I,\omega^I$  $(I=1,\cdots,6)$, transforming as the adjoint
representation of $SU(2)^6$.
A model in this construction
is defined by a set of boundary condition basis vectors,
and the one--loop GSO projection coefficients.
The basis vectors generate a finite additive group
$\Xi$. The physical states in the Hilbert space, of a given sector
$\alpha\in\Xi$, are obtained by acting on the vacuum
with bosonic and fermionic operators. For a periodic complex fermion $f$,
there are two degenerate vacua ${\vert +\rangle},{\vert -\rangle}$ ,
annihilated by the zero modes $f_0$ and
${{f_0}^*}$ and with fermion numbers  $F(f)=0,-1$, respectively.
The physical spectrum is obtained by applying
the generalized GSO projections.

The free fermion three generation models
are constructed in two stages. The first
corresponds to the NAHE set of boundary basis vectors
$\{{\bf1},S,b_1,b_2,b_3\}$ \cite{nahe,rffm}. The second consists
of adding to the NAHE set three additional boundary
condition basis vectors, typically denoted $\{\alpha,\beta,\gamma\}$.
The sector ${S}$ generates $N=4$ space--time supersymmetry,
which is broken to $N=2$ and $N=1$ space--time supersymmetry by
$b_1$ and $b_2$, respectively. The gauge group after the NAHE set is
$SO(10)\times E_8\times SO(6)^3$, 
which is broken to $SO(4)^3\times U(1)^3\times
SO(10)\times SO(16)$ by the vector $2\gamma$.
At the level of the NAHE set, each
sector $b_1$, $b_2$ and  $b_3$  give rise to 16 spinorial 16 of $SO(10)$.
The Neveu-Schwarz (NS) sector produces some massless states that transform as
$(5\oplus\bar5)$ of $SO(10)$ and some others that are singlets of $SO(10)$. 
All the states from the NS sector are singlets of the hidden $E_8$.

The NAHE set divides the internal world--sheet fermions into several
groups.
The internal $44$ right--moving  fermionic states
are divided in the following way:
${\bar\psi}^{1,\cdots,5}$ are complex and produce the observable $SO(10)$
symmetry;
${\bar\phi}^{1,\cdots,8}$ are complex and produce the hidden $E_8$ gauge
group;
$\{{\bar\eta}^1,{\bar y}^{3,\cdots,6}\}$, $\{{\bar\eta}^2,{\bar y}^{1,2}
,{\bar\omega}^{5,6}\}$, $\{{\bar\eta}^3,{\bar\omega}^{1,\cdots,4}\}$
 give rise to the three horizontal $SO(6)$ symmetries.
The left--moving $\{y,\omega\}$ states
are divided to,
$\{{y}^{3,\cdots,6}\}$, $\{{y}^{1,2}
,{\omega}^{5,6}\}$, $\{{\omega}^{1,\cdots,4}\}$.
The left--moving $\chi^{12},\chi^{34},\chi^{56}$ states
carry the supersymmetry charges.

An important consequence of the NAHE set is
observed by extending the $SO(10)$
symmetry to $E_6$. Adding
to the NAHE set a vector $\xi_2$
with periodic boundary conditions for the set
$\{{{\bar\psi}^{1,\cdots,5}},
{{\bar\eta}^{1,2,3}}\}$, extends the gauge symmetry to
$E_6\times U(1)^2\times SO(4)^3$.
Each spinorial 16 of $SO(10)$, produced by one of the three
sectors $b_j$, combines with a 10+1 of $SO(10)$, produced by
the sector $b_j+\xi_2$, to give a 27 of $E_6$.
The sectors $(b_j;b_j+\xi_2)$, $(j=1,2,3)$
each give eight $27$ of $E_6$. The untwisted $(NS;NS+\xi_2)$ sector gives, in
addition to the vector bosons and spin two states, three copies of
scalar representations in $27+{\bar {27}}$ of $E_6$.
Alternatively,
we can start with an extended NAHE set $\{{\bf1},S,\xi_1,\xi_2,b_1,
b_2\}$, with $\xi_1={\bf1}+b_1+b_2+b_3$. The set $\{{\bf1},S,\xi_1,
\xi_2\}$ produces a toroidal Narain model with $SO(12)\times
E_8\times E_8$ or $SO(12)\times SO(16)\times SO(16)$ gauge
group depending on the GSO phase $c({\xi_1\atop\xi_2})$.
The basis vectors $b_1$ and $b_2$ then break $SO(12)\rightarrow
SO(4)^3$, and either $E_8\times E_8\rightarrow E_6\times U(1)^2\times E_8$
or $SO(16)\times SO(16)\rightarrow SO(10)\times U(1)^3\times SO(16)$.
The vectors $b_1$ and $b_2$ correspond to ${Z}_2\times {Z}_2$ orbifold
modding. The three sectors $b_1$, $b_2$ and $b_3$ correspond to
the three twisted sectors of the ${Z}_2\times {Z}_2$ orbifold,
with each producing eight generations in the ${\bf27}$, or
${\bf16}$, representations
of $E_6$, or $SO(10)$, respectively. In the case of $E_6$ the untwisted
sector produces an additional $3\times ({\bf 27}+{\overline{\bf 27}})$,
whereas in the $SO(10)$ model
it produces $3\times({\bf 10}+{\overline{\bf 10}})$.
Therefore, the Calabi--Yau manifold that corresponds to the
${Z}_2\times {Z}_2$ orbifold at the free fermionic point
in the Narain moduli space has $(h_{11},h_{21})=(27,3)$.

In this model the fermionic states which count the
multiplets of $E_6$ are the internal fermions $\{y,w\vert{\bar y},
{\bar\omega}\}$. The vacuum of the sectors
$b_j$  contains twelve periodic fermions. Each periodic fermion
gives rise to a two dimensional degenerate vacuum $\vert{+}\rangle$ and
$\vert{-}\rangle$ with fermion numbers $0$ and $-1$, respectively.
After applying the
GSO projections, we can write the vacuum of the sector
$b_1$ in combinatorial form
\begin{eqnarray}
{\left[\left(\matrix{4\cr
                                    0\cr}\right)+
\left(\matrix{4\cr
                                    2\cr}\right)+
\left(\matrix{4\cr
                                    4\cr}\right)\right]
\left\{\left(\matrix{2\cr
                                    0\cr}\right)\right.}
&{\left[\left(\matrix{5\cr
                                    0\cr}\right)+
\left(\matrix{5\cr
                                    2\cr}\right)+
\left(\matrix{5\cr
                                    4\cr}\right)\right]
\left(\matrix{1\cr
                                    0\cr}\right)}\nonumber\\
{+\left(\matrix{2\cr
                                    2\cr}\right)}
&{~\left[\left(\matrix{5\cr
                                    1\cr}\right)+
\left(\matrix{5\cr
                                    3\cr}\right)+
\left(\matrix{5\cr
                                    5\cr}\right)\right]\left.
\left(\matrix{1\cr
                                    1\cr}\right)\right\}}
\label{spinor}
\end{eqnarray}
where
$4=\{y^3y^4,y^5y^6,{\bar y}^3{\bar y}^4,
{\bar y}^5{\bar y}^6\}$, $2=\{\psi^\mu,\chi^{12}\}$,
$5=\{{\bar\psi}^{1,\cdots,5}\}$ and $1=\{{\bar\eta}^1\}$.
The combinatorial factor counts the number of $\vert{-}\rangle$ in
a given state. The two terms in the curly brackets correspond to the two
components of a Weyl spinor.  The $10+1$ in the $27$ of $E_6$ are
obtained from the sector $b_j+\xi_1$.
The states which count the multiplicities of $E_6$ are the internal
fermionic states $\{y^{3,\cdots,6}\vert{\bar y}^{3,\cdots,6}\}$.
A similar result is
obtained for the sectors $b_2$ and $b_3$ with $\{y^{1,2},\omega^{5,6}
\vert{\bar y}^{1,2},{\bar\omega}^{5,6}\}$
and $\{\omega^{1,\cdots,4}\vert{\bar\omega}^{1,\cdots,4}\}$
respectively, which suggests that
these twelve states correspond to a six dimensional
compactified orbifold with Euler characteristic equal to 48.

The construction of the free fermionic models beyond the NAHE--set
entails the construction of additional boundary condition basis 
vectors and the associated one--loop GSO phases. Their
function is to reduce the number of generations and 
at the same time break the four dimensional gauge group.
In terms of the former, the reduction is primarily by the
action on the set of internal world--sheet fermions
$\{y,\omega|{\bar y},{\bar\omega}\}$.
As elaborated in the next section this 
set corresponds to the internal compactified manifold
and the action of the additional boundary condition basis
vectors on this set also breaks the gauge symmetries 
from the internal lattice enhancement. The latter is 
obtained by the action on the gauge degrees of freedom
which correspond to the world--sheet fermions
$\{{\bar\psi}^{1,\cdots,5},{\bar\eta}^{1,\cdots,3},
{\bar\phi}^{1,\cdots,8}\}$.
In the bosonic formulation this would correspond to
Wilson--line breaking of the gauge symmetries, hence for the purpose
of the reduction of the number of generations we
can focus on the assignment to the internal world--sheet fermions
$\{y,\omega|{\bar y},{\bar\omega}\}$.

We can therefore examine basis vectors that do not break the
gauge symmetries further, {\it i.e.} basis vectors of the 
form $b_j$, with $\{\psi^\mu_{1,2}\chi_{j,j+1},(y,\omega|{\bar y},
{\bar\omega}),{\bar\psi}^{1,\cdots,5},{\bar\eta}_j\}=1$
for some selection of $(y,\omega|{\bar y},{\bar\omega})=1$
assignments such that the additional vectors
$b_j$ produce massless $SO(10)$ spinorials. We will refer
to such vectors as spinorial vectors. The additional
basis vectors $b_j$ can then produce chiral, or non--chiral, spectrum. 
The condition that the spectrum from a given such sector $b_j$
be chiral is that there exist another spinorial
vector, $b_i$, in the additive
group $\Xi$, such that the overlap between the
periodic fermions of the internal set $(y,\omega|{\bar y},
{\bar\omega})$ is empty, {\it i.e.}
\beq
\{b_j(y,\omega|{\bar y},{\bar\omega})\}\cap
\{b_i(y,\omega|{\bar y},{\bar\omega})\}\equiv\emptyset~.
\label{chiralitycondition}
\eeq
If there exists such a vector $b_i$ in the additive group then it
will induce a GSO projection that will select
the chiral states from the sector $b_j$. Interchangeably,
if such a vector does not exist, states from the sector
$b_j$ will be non--chiral, ${\it i.e.}$ there will be an equal
number of $16$ and $\overline{16}$ or $27$ and $\overline{27}$.
For example, we note that for the NAHE--set basis vectors
the condition (\ref{chiralitycondition}) is satisfied.
In section (\ref{z2z2corres}) we will discuss the geometrical
correspondence of this condition.
The reduction to three generations in a specific model
is illustrated in table \ref{m278}.

\beqn
 &\begin{tabular}{c|c|ccc|c|ccc|c}
 ~ & $\psi^\mu$ & $\chi^{12}$ & $\chi^{34}$ & $\chi^{56}$ &
        $\bar{\psi}^{1,...,5} $ &
        $\bar{\eta}^1 $&
        $\bar{\eta}^2 $&
        $\bar{\eta}^3 $&
        $\bar{\phi}^{1,...,8} $\\
\hline
\hline
  ${\alpha}$  &  0 & 0&0&0 & 1~1~1~0~0 & 0 & 0 & 0 & 1~1~1~1~0~0~0~0 \\
  ${\beta}$   &  0 & 0&0&0 & 1~1~1~0~0 & 0 & 0 & 0 & 1~1~1~1~0~0~0~0 \\
  ${\gamma}$  &  0 & 0&0&0 &
		${1\over2}$~${1\over2}$~${1\over2}$~${1\over2}$~${1\over2}$
	      & ${1\over2}$ & ${1\over2}$ & ${1\over2}$ &
                ${1\over2}$~0~1~1~${1\over2}$~${1\over2}$~${1\over2}$~0 \\
\end{tabular}
   \nonumber\\
   ~  &  ~ \nonumber\\
   ~  &  ~ \nonumber\\
     &\begin{tabular}{c|c|c|c}
 ~&   $y^3{y}^6$
      $y^4{\bar y}^4$
      $y^5{\bar y}^5$
      ${\bar y}^3{\bar y}^6$
  &   $y^1{\omega}^5$
      $y^2{\bar y}^2$
      $\omega^6{\bar\omega}^6$
      ${\bar y}^1{\bar\omega}^5$
  &   $\omega^2{\omega}^4$
      $\omega^1{\bar\omega}^1$
      $\omega^3{\bar\omega}^3$
      ${\bar\omega}^2{\bar\omega}^4$ \\
\hline
\hline
$\alpha$ & 1 ~~~ 0 ~~~ 0 ~~~ 0  & 0 ~~~ 0 ~~~ 1 ~~~ 1  & 0 ~~~ 0 ~~~ 1 ~~~ 1
\\
$\beta$  & 0 ~~~ 0 ~~~ 1 ~~~ 1  & 1 ~~~ 0 ~~~ 0 ~~~ 0  & 0 ~~~ 1 ~~~ 0 ~~~ 1
\\
$\gamma$ & 0 ~~~ 1 ~~~ 0 ~~~ 1  & 0 ~~~ 1 ~~~ 0 ~~~ 1  & 1 ~~~ 0 ~~~ 0 ~~~ 0
\\
\end{tabular}
\label{m278}
\eeqn
In the realistic free fermionic models the vector $X$
is replaced by the vector $2\gamma$ in which $\{{\bar\psi}^{1,\cdots,5},
{\bar\eta}^1,{\bar\eta}^2,{\bar\eta}^3,{\bar\phi}^{1,\cdots,4}\}$
are periodic. This reflects the fact that these models
have (2,0) rather than (2,2) world-sheet supersymmetry.
At the level of the NAHE set we have 48 generations.
One half of the generations is projected because of the vector $2\gamma$.
Each of the three vectors in table \ref{m278}
acts nontrivially on the degenerate
vacuum of the fermionic states
$\{y,\omega\vert{\bar y},{\bar\omega}\}$ that are periodic in the
sectors $b_1$, $b_2$ and $b_3$ and reduces the combinatorial
factor of Eq. (\ref{spinor}) by a half.
Thus, we obtain one generation from each sector $b_1$, $b_2$ and $b_3$.
By replacing the basis vectors $\alpha$, $\beta$, $\gamma$
with $\alpha+\beta$, $\alpha+\gamma$ and $\alpha+\beta+\gamma$,
it is seen that the action on the internal coordinates
of two of the basis vectors beyond the NAHE--set correspond 
to symmetric shifts, whereas the third corresponds to a fully asymmetric 
shift \cite{fknr}. In sections
(\ref{lowrankorbifolds}--\ref{beyondz2z2}) we will classify
all the possible symmetric shifts on $Z_2\times Z_2$ orbifolds.

\setcounter{footnote}{0}
\section{The $Z_2\times Z_2$ correspondence}\label{z2z2corres}

In this section we elaborate on the correspondence between the free fermion
models and $Z_2\times Z_2$ orbifold. The aim is to set the stage for
the analysis of the $Z_2\times Z_2$ orbifold beyond the NAHE--set
correspondence. In this respect we remark that the NAHE--set
is a particular realization of the $Z_2\times Z_2$ orbifold 
at a fixed point in the moduli space. However, its crucial property
is precisely its correspondence with a $Z_2\times Z_2$ orbifold. 
For many issues pertaining to the phenomenology of the
relevant string vacua, it will prove beneficial to abandon
the free fermionic realization and to resort to the bosonic,
or geometrical description. In this respect the NAHE based
free fermionic models merely indicate that the relevant geometrical 
structure for this class of models is that of the
$Z_2\times Z_2$ orbifold.

To translate the fermionic boundary conditions to twists and shifts in the 
bosonic formulation we bosonize the real fermionic degrees of freedom, 
$\{y,\omega\vert{\bar y},{\bar\omega}\}$. Defining, 
${\xi_i}={\sqrt{1\over2}}(y_i+i\omega_i)=-ie^{iX_i}$, 
$\eta_i={\sqrt{1\over2}}(y_i-i\omega_i)=-ie^{-iX_i}$
with similar definitions for the right movers $\{{\bar y},{\bar\omega}\}$
and $X^I(z,{\bar z})=X^I_L(z)+X^I_R({\bar z})$. 
With these definitions the world--sheet supercurrents in the bosonic 
and fermionic formulations are equivalent,
$T_F^{int}=\sum_{i}\chi_i{y_i}\omega_i=i\sum_{i}\chi_i{\xi_i}\eta_i=
\sum_{i}\chi_i\partial{X_i}.$ The momenta $P^I$ of the compactified scalars
in the bosonic formulation are identical with the $U(1)$ charges 
$Q(f)$ of the unbroken Cartan generators of the four dimensional
gauge group, 
$$Q(f)={1\over2}\alpha(f)+F(f)$$
where $\alpha(f)$ are the boundary conditions of complex fermions $f$,
reduced to the interval $(-1,1]$ and $F(f)$ is a fermion number operator. 

The extended NAHE--set model is generated in the
orbifold language by modding out an $SO(12)$ lattice by a $Z_2\times{Z_2}$
discrete symmetry with standard embedding \cite{foc}. The $SO(12)$ lattice
is obtained for special values of the metric and antisymmetric tensor
and at a fixed point in compactification space. The metric is the Cartan
matrix of $SO(12)$ and the antisymmetric tensor is given by $b_{ij}=g_{ij}$
for $i>j$.
The boundary condition vectors $b_1$ and $b_2$ translate into 
$Z_2\times Z_2$ twists on the bosons $X_i$ and fermions $\chi_i$ and 
to shifts on the gauge degrees of freedom. The massless spectrum
of the resulting orbifold model consist of the untwisted sector and 
three twisted sectors, $b_1$, $b_2$ and $b_3$.
Starting from the Narain model with $SO(12)\times E_8\times E_8$
symmetry~\cite{Narain}, and applying the $Z_2\times Z_2$ twisting on the
internal
coordinates, we then obtain the orbifold model with $SO(4)^3\times
E_6\times U(1)^2\times E_8$ gauge symmetry. There are eight fixed
points in each twisted sector, yielding the 24 generations from the
three twisted sectors. The three additional pairs of $27$
and $\overline{27}$ are obtained from the untwisted sector. This
orbifold model exactly corresponds to the free-fermion model
with the six-dimensional basis set
$\{{\bf1},S,\xi_1,\xi_2,b_1,b_2\}$.
The Euler characteristic of this model is 48 with $h_{11}=27$ and
$h_{21}=3$.
We refer to this model as $X_2$.

This $Z_2\times Z_2$ orbifold, corresponding
to the extended NAHE set, at the core of the realistic
free fermionic models, differs from the one
at a generic point in the moduli space. 
In that $Z_2\times Z_2$ orbifold model
the Euler characteristic is equal to 96,
or 48 generations, and $h_{11}=51$, $h_{21}=3$.
We refer to this model as $X_1$.

For the purpose of the analysis in section \ref{beyondz2z2}
it is instructive to discuss the connection between the
$X_1$ and $X_2$ models. We consider here only
symmetric shifts on the toroidal coordinates.
We start by constructing
the ${Z}_2\times {Z}_2$ at a generic point in the
moduli space. The compactified
$T^2_1\times T^2_2\times T^2_3$ torus is parameterized by
three complex coordinates $z_1$, $z_2$ and $z_3$,
with the identification
\begin{equation}
z_i=z_i + 1~~~~~~~~~~;~~~~~~~~~~z_i=z_i+\tau_i,
\label{t2cube}
\end{equation}
where $\tau$ is the complex parameter of each torus
$T^2$.
We consider ${Z}_2$ twists and possible shifts of order
two:
\begin{equation}
z_i~\rightarrow~(-1)^{\epsilon_i}z_i+{1\over 2}\delta_i,
\label{z2twistanddance}
\end{equation}
subject to the condition that $\Pi_i(-1)^{\epsilon_i}=1$.
This condition insures that the holomorphic three--form
$\omega=dz_1\wedge dz_2\wedge dz_3$ is invariant under the ${Z}_2$ twist.
Under the identification $z_i\rightarrow-z_i$, a single torus
has four fixed points at
\begin{equation}
z_i=\{0,1/2,\tau/2,(1+\tau)/2\}.
\label{fixedtau}
\end{equation}
The first model that we consider is produced
by the two ${Z}_2$ twists
\begin{eqnarray}
&& \alpha:(z_1,z_2,z_3)\rightarrow(-z_1,-z_2,~~z_3)\cr
&&  \beta:(z_1,z_2,z_3)\rightarrow(~~z_1,-z_2,-z_3).
\label{alphabeta}
\end{eqnarray}
There are three twisted sectors in this model, $\alpha$,
$\beta$ and $\alpha\beta=\alpha\cdot\beta$, each producing
16 fixed tori, for a total of 48. The untwisted sector
adds three additional K\"ahler and complex deformation
parameters producing in total a manifold with $(h_{11},h_{21})=(51,3)$.

Next we consider the model generated by the ${Z}_2\times {Z}_2$
twists in (\ref{alphabeta}), with the additional shift
\begin{equation}
\gamma:(z_1,z_2,z_3)\rightarrow(z_1+{1\over2},z_2+{1\over2},z_3+{1\over2}).
\label{gammashift}
\end{equation}
This model again has fixed tori from the three
twisted sectors $\alpha$, $\beta$ and $\alpha\beta$.
The product of the $\gamma$ shift in (\ref{gammashift})
with any of the twisted sectors does not produce any additional
fixed tori. Therefore, this shift acts freely.
Under the action of the $\gamma$ shift, half
the fixed tori from each twisted sector are paired.
Therefore, the action of this shift is to reduce
the total number of fixed tori from the twisted sectors
by a factor of $1/2$. Consequently, in this model
$(h_{11},h_{21})=(27,3)$. This model therefore
reproduces the data of the ${Z}_2\times {Z}_2$ orbifold
at the free-fermion point in the Narain moduli space.

To facilitate the
discussion of the subsequent examples, we briefly describe
the calculation of the cohomology for this orbifold:
a more complete discussion can be found in~\cite{msanddt}.
Consider first the untwisted sector. The Hodge
diamond for a single untwisted torus is given by
\beq
\left(\matrix{1 &1\cr
              1 &1\cr
             } \right)
\label{hodge1t}
\eeq
which displays the dimensions of the $H^{p,q}(T_i)$,
with $H^{0,0}$, $H^{0,1}$, $H^{1,0}$ and $H^{1,1}$
being generated by the differential forms $1$, $d{\bar z}_i$,
$dz_i$ and $dz_i\wedge d{\bar z}i$, respectively. Under the $Z_2$
transformation $z~\rightarrow~-z$, $H^{0,0}$ and $H^{1,1}$ 
are invariant, whereas $H^{1,0}$ and $H^{0,1}$ change sign.

The untwisted sector of the manifold produced
by the product of the three tori $T_1\times T_2\times T_3$
is then given by the product of differential forms
which are invariant under the $Z_2\times Z_2$ twists
$\alpha\times \beta$. The invariant terms
are summarized by the Hodge diamond
\beq
\left(\matrix{ 1&0&0&1\cr
               0&3&3&0\cr
               0&3&3&0\cr
               1&0&0&1\cr}\right)
\label{hodge3t}
\eeq
For example, $H^{1,1}$ is generated by $dz_i\wedge {\bar z}_i$ 
for $i=1,2,3$, and $H^{2,1}$ is produced by 
$dz_1\wedge z_2\wedge{\bar z}_3$, $dz_2\wedge z_3\wedge{\bar z}_1$,
$dz_3\wedge z_1\wedge{\bar z}_2$, etc..
We next turn to the twisted sectors, of which there are three,
produced by $\alpha$, $\beta$ and $\alpha\beta$, respectively.
In each sector, two of the $z_i$ are identified under $z_i\rightarrow-z_i$,
and one torus is left invariant. We need then consider
only one of the twisted sectors, say $\alpha$, and the others
will contribute similarly. The sector $\alpha$ has 16 fixed
points from the action of the twist on the first and second
tori. Since the action is trivial on the third torus, we get 16 fixed
tori. The cohomology
is given by sixteen copies of the cohomology of $T_3$,
where each $H^{p,q}$ of $T_3$ contributes $H^{p+1,q+1}$
to that of the orbifold theory~\cite{msanddt}.
The Hodge diamond from each twisted sector then 
has the form
\beq
\left(\matrix{ 0& 0& 0&0\cr
               0&16&16&0\cr
               0&16&16&0\cr
               0& 0& 0&0\cr}\right)
\label{hodgets}
\eeq
It remains to find the forms from the $\alpha$ twisted sector
which are invariant under the action of the $\beta$ twist. 
Since $z_3\rightarrow-z_3$ under $\beta$, it follows that 
$1$ and $dz_3\wedge d{\bar z}_3$ are invariant,
whereas $dz_3$ and $d{\bar z}_3$ are not.
Consequently, only the contributions of $H^{1,1}$
and $H^{2,2}$ in (\ref{hodgets}) are invariant 
under the $\beta$ twist. 
Therefore, we see that the invariant contribution
from each twisted sector is only along the diagonal of
(\ref{hodgets}), and that the total Hodge diamond of the
$Z_2\times Z_2$ orbifold is 
\beq
\left(\matrix{ 1& 0& 0&1\cr
               0&51& 3&0\cr
               0& 3&51&0\cr
               1& 0& 0&1\cr}\right)
\label{hodgez2z2}
\eeq

Next we consider the model generated by the $Z_2\times Z_2$
twists in (\ref{alphabeta}), with the additional shift Eq. (\ref{gammashift}).
This model again has fixed tori from the three
twisted sectors $\alpha$, $\beta$ and $\alpha\beta$.
The product of the $\gamma$ shift in (\ref{gammashift})
with any of the twisted sectors does not produce any additional
fixed tori. Therefore, this shift acts freely.
Under the action of the $\gamma$ shift, half
the fixed tori from each twisted sector are paired.
Therefore, the action of this shift is to reduce
the total number of fixed tori from the twisted sectors
by a factor of $1/2$. Consequently, the Hodge diamond for this model is
\beq
\left(\matrix{ 1& 0& 0&1\cr
               0&27& 3&0\cr
               0& 3&27&0\cr
               1& 0& 0&1\cr}\right)
\label{hodgez2z2shift3}
\eeq
with $(h_{11},h_{21})=(27,3)$. This model therefore
reproduces the data of the $Z_2\times Z_2$ orbifold
at the free-fermion point in the Narain moduli space.

Finally, let us consider the model generated by the
twists (\ref{alphabeta}) with the additional
shift given by
\beq
\gamma:(z_1,z_2,z_3)\rightarrow(z_1+{1\over2},z_2+{1\over2},z_3)
\label{gammashift2}
\eeq
This model, denoted by $X_3$, again has three twisted sectors $\alpha$,
$\beta$ and $\alpha\beta$. Under the action of the $\gamma$ shift,
half of the fixed tori from these twisted
sectors are identified. These twisted sectors
therefore contribute to the Hodge diamond 
as in the previous model. However, the $\gamma$ shift
in (\ref{gammashift2}) does not act freely,
as its combination with $\alpha$ produces additional 
fixed tori, since, under the action of the product
$\alpha\cdot\gamma$, we have
\beq
\alpha\gamma:(z_1,z_2,z_3)\rightarrow(-z_1+{1\over2},-z_2+{1\over2},z_3)
\label{alphagamma}
\eeq
This sector therefore has 16 additional fixed tori.
Repeating the analysis as in the previous cases,
we see that, under the identification imposed by the $\alpha$ and $\beta$ 
twists, the invariant states from this sector
give rise to four additional (1,1) forms and four additional
(2,1) forms. The Hodge diamond for this model
therefore has the form
\beq
\left(\matrix{ 1& 0& 0&1\cr
               0&31& 7&0\cr
               0& 7&31&0\cr
               1& 0& 0&1\cr}\right)
\label{hodgez2z2shift4}
\eeq
with $(h_{11},h_{21})=(31,7)$.

String theory allows
more complicated operations that involve shifts of momentum, 
winding, or both \cite{vwaaf}.
Whereas the first is symmetric between left and right
movers, the last is asymmetric.
In fact, it was shown in \cite{foc} that the $SO(12)$ lattice at
the free fermionic point is reproduced by such an asymmetric shift,
that differs from (\ref{gammashift}). These 
different freely acting shifts induce the same projection
on the massless spectrum, but the partition function and the massive
spectrum differ.
We can regard all the resulting
quotient manifolds as existing in the same moduli space, that
may be connected by continuous extrapolations.
In this paper we restrict the analysis to symmetric
shifts, in the spirit that the free fermionic models 
merely reveal the central role of the $Z_2\times Z_2$
orbifold. Thus, our aim is to promote the understanding
of the geometry of the $Z_2\times Z_2$ orbifolds, and
in particular with respect to the phenomenological
features exhibited by the free fermionic models.
Namely, in respect to the fashion in which the three massless
chiral generations arise in the free fermionic models. 

In this regard, in the orbifold picture
we start with the (51,3) $Z_2\times Z_2$ orbifold. 
We can regard each twisted sector as generated by a
two coordinate base and one coordinate fiber, with
the twists acting on the coordinates of the base.
Above each of the 16 fixed points we then have
the untwisted fiber which is an unfixed torus. 
We can then imagine that in this picture the
reduction to three generations entails the reduction
of the number of independent fixed point, by identifying 
points on the base by shifts. Indeed, the action of
(\ref{gammashift}) is precisely such a reduction of the
number of twisted fixed points from 48 to 24. Thus, 
we can imagine imposing additional shift operations that
will reduce the number of fixed points further. From
the analysis of (\ref{gammashift}) and (\ref{gammashift2})
we note that the additional shifts can be freely acting
or non--freely acting. In the following we perform 
a complete classifications of all possible symmetric
shifts and prove that a reduction to
three fixed points is not possible solely with symmetric 
shifts. This result concurs with the free fermionic
analysis that indicates that at least one asymmetric
operation is required to reduce the number of families
to three generations \cite{fknr}.

\section{Classification of low rank orbifolds}\label{lowrankorbifolds}
We are interested in orbifolds of the form 
\begin{equation}
X=(E_1\times E_2\times E_3)/G
\label{Xg}
\end{equation}
Here the three $E_i$ are elliptic curves, or topologically they
are tori $T^2$. The group $G$ contains the $(Z_2)^2$ of twists,
as well as a subgroup ${\overline G}$, which acts on
$E_1\times E_2\times E_3$ by translations. In this work
we consider only translations of order 2. This means that
our ${\overline G}$ is a subgroup of the group
\begin{equation}
E_1[2]\times E_2[2]\times E_3[2]\approx (Z_2)^6
\label{subgroup}
\end{equation}
of points of order 2 on the elliptic curves. 
Here $E_i[2]$ is the group
$(Z_2)^2$ of points of order 2 in $E_i$. 
We denote its four elements by: $0,~1,~\tau,~1+\tau$ (they should 
be more accurately labeled as $1/2~,\tau/2,~(1+\tau)/2$, but the 
notation would then get out of hand). The full group $G$ is
an extension 
\begin{equation}
0\rightarrow (Z_2)^2\rightarrow G\rightarrow {\overline G}\rightarrow 0,
\label{gextension}
\end{equation}
{\it i.e.} it contains $(Z_2)^2$ as a subgroup and the
quotient $G/(Z_2)^2$ is ${\overline G}$. In general,
this extension can be non--trivial. However, in our case
of translations of order 2, the entire $G$ is commutative,
so the extension is really a product,
\beq
G\approx (Z_2)^2\times {\overline G}.  
\label{entireg}
\eeq
The obvious invariant of a group ${\overline G}$
is its rank $r$, {\it i.e.} the number of its generators
(or its dimension as a vector space over $Z_2$).
The set of all subgroups ${\overline G}\subset (Z_2)^6$ of
a given rank $r$ is called a (finite)
Grassmannian $Gr_{Z_2}(r,6)$. The number
of subgroups of rank $r$ is given by: 
\beq
\prod_{i=1}^r{{64-2^{(i-1)}}\over{2^r-2^{(i-1)}}}.
\eeq
This is seen by noting that the first generator
of ${\overline G}$ can be any of the 63 non--zero
elements. Having chosen it, the second generator
cannot be a multiple, so has only $62=64-2$
options. The third generator needs to avoid 
the four elements, which are combinations
of the first two, hence $60=64-4$ options,
and so on. But each subgroup has now been counted
as many times as we can choose bases for it. 
The choice of a basis amounts to $2^r-1$ possibilities
for the first vector, $2^r-2$ for the second, and so on.

Explicitly, these numbers are 
\beq
 \begin{tabular}{c|ccccccc}
r & 0 & 1 & 2 & 3  & 4 & 5 & 6\\
\hline
\# & 1 & 63&651&1395&651&63& 1\\
\end{tabular}
\eeq
Fortunately, we do not need to consider that many
cases. The permutation group $S_3$ acts on the 3
non--zero elements of each $E_i[2]$. This gives a
total symmetry group of $(S_3)^3$. In addition, one 
more copy of $S_3$ acts on the product 
$E_1[2]\times E_2[2]\times E_3[2]$ by
permuting the 3 tori. The full symmetry group
$S$ is therefore an extension of $S_3$ by $(S_3)^3$:
\beq
0\rightarrow (S_3)^3\rightarrow S\rightarrow S_3\rightarrow 0.
\eeq
A group of rank 1 is uniquely specified by a single
non--zero element $(a_1,a_2,a_3)\in E_1[2]\times E_2[2]\times E_3[2]$.
The 63 original possibilities are reduced by $S$
to just 3 equivalence classes, namely those of, 
\beqn
        & (1,1,1); \nonumber\\
        & (1,1,0); \nonumber\\
        & (1,0,0).
\label{genericform1}
\eeqn
Indeed, we use the overall $S_3$ to move the non--zero $a_i$
to the left and the zeroes to the right. Then we use the
individual symmetries $S_3^i$ of the $E_i$ to change each
non--zero entry to a $1$, resulting in just the above
three classes. Note that the first of these groups
corresponds to the $(27,3)$ model, denoted $X_2$ in section \ref{z2z2corres}.
The (51,3) model, denoted $X_1$ in section \ref{z2z2corres}, corresponds 
to the unique ${\overline G}$ of rank 0. The Hodge numbers
of the other two groups listed above can be computed 
according to rules explained in section 6 below. They are
(31,7) for (1,1,0) and (51,3) for (1,0,0).

With a bit of patience, one can similarly work out the complete
list of rank 2 subgroups: 
\beq
 \begin{tabular}{|c|c|ccc|c|}
\hline
(1,1,1) & ($\tau ,\tau ,\tau$ ) & 0 & 0 & 0 & (15,3)\\
(1,1,1) & (0     ,$\tau,\tau$ ) & 0 & 1 & 0 & (17,5)\\
(1,1,1) & (0     ,    0,$\tau$) & 0 & 2 & 0 & (27,3)\\
(1,1,1) & ($\tau$,    1,   0  ) & 0 & 1 & 1 & (19,7)\\
(1,1,1) & (1     ,    0,   0  ) & 0 & 2 & 1 & (27,3)\\
(1,1,0) & ($\tau ,\tau$,   0  ) & 1 & 1 & 1 & (21,9)\\
(1,1,0) & (1     ,    0,   1  ) & 1 & 1 & 1 & (27,3)\\
(1,0,0) & (0     ,    1,   0  ) & 2 & 2 & 1 & (51,3)\\
(1,0,0) & ($\tau$,    1,   0  ) & 2 & 1 & 1 & (31,7)\\
(1,0,0) & ($\tau$,    0,   0  ) & 2 & 2 & 2 & (51,3)\\
\hline
\end{tabular}
\eeq
Next to each group we itemized the number of zero
entries in each of its three non--zero elements.
There is only one pair (the $6th$ and $7th$ groups)
with the same distribution of zeroes, but these two
are easily seen to be non--equivalent anyway ({\it e.g.}
because the zeroes are always in the third entry for group 
\# 6, but in varying entries for \# 7). In the last column we listed
the Hodge numbers $(h^{11},h^{21})$ for each orbifold
$T^6/G$. The rules for calculating these Hodge
numbers are explained in section \ref{rulesfororco} below. 

\section{Classification of rank 3 orbifolds}\label{section5}
A subgroup ${\overline G}$ of rank 3 contains 7 non--zero
elements and 7 rank 2 subgroups. Each rank 2 subgroup
contains 3 non--zero elements, and each non--zero 
element belongs to 3 rank 2 subgroups. We can display
the situation by the following diagram
\begin{figure}[!h]
\centerline{\epsfxsize 1.0 truein \epsfbox{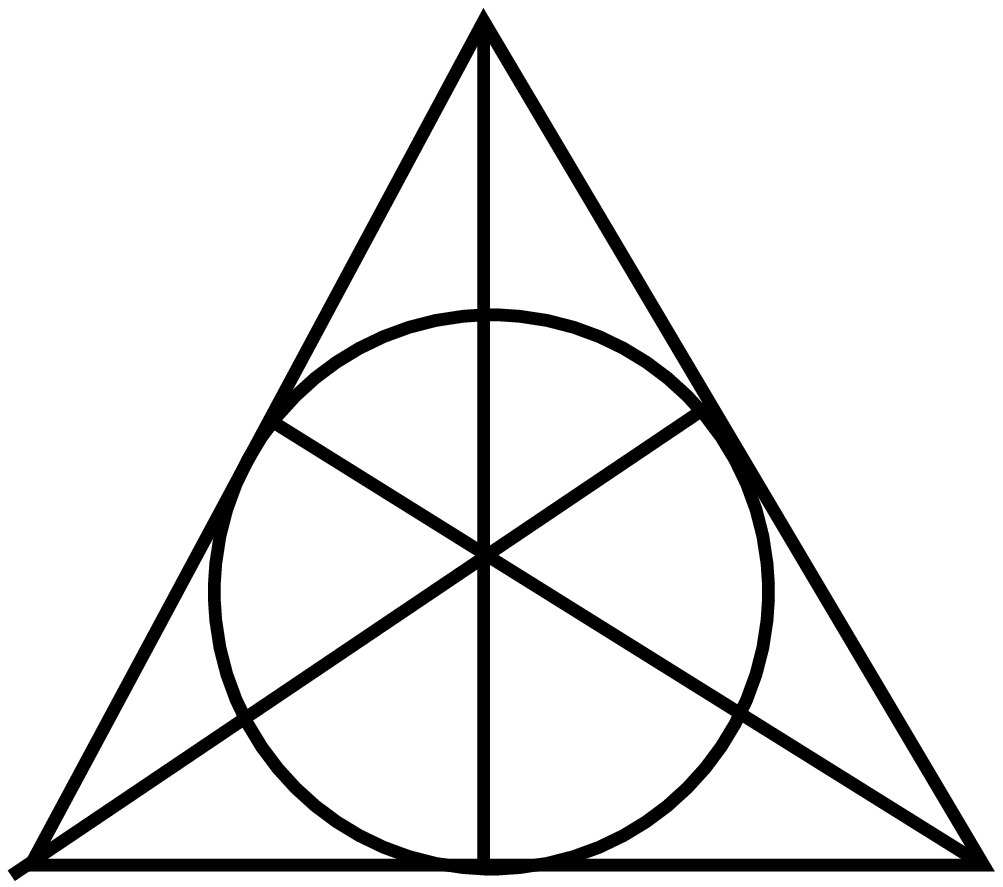}}
\label{figure1}
\end{figure}

The vertices represent non--zero elements while
the 6 edges and the circle represent rank 2 subgroups.
To each rank 3 group ${\overline G}$ we assign a decorated diagram,
{\it e.g.} for the group
${\overline G}:~(1,1,1),~(1,0,0),~(\tau,0,0)$ the decorated
diagram is:
\begin{figure}[!h]
\centerline{\epsfxsize 1.0 truein \epsfbox{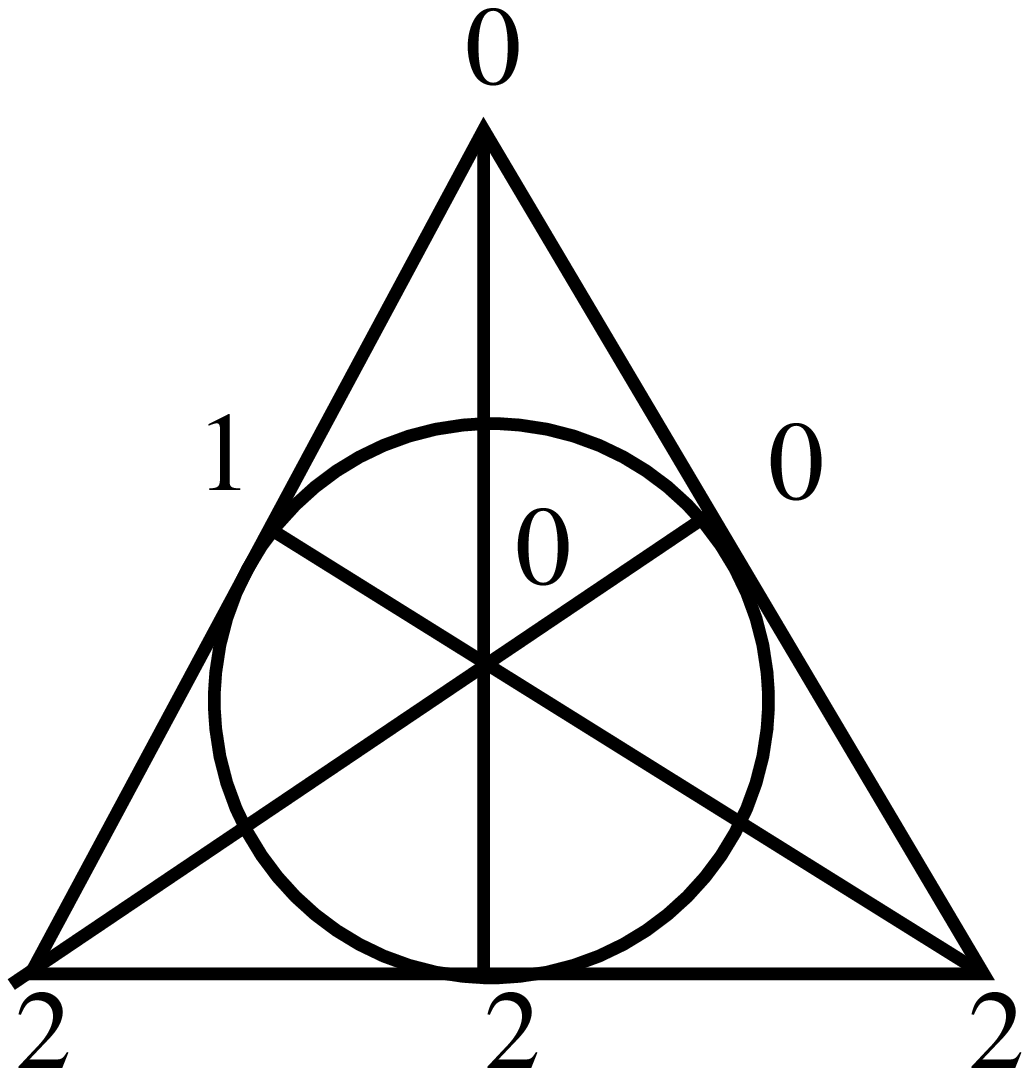}}
\label{figure2}
\end{figure}

The 0 on the top vertex gives the number of zero entries
in the first generator $(1,1,1)$. The bottom left and right
vertices correspond to those in $(1,0,0)$ and $(\tau,0,0)$. The entries
on the edges correspond similarly to the other group elements. 

\begin{figure}[!h]
\centerline{\epsfxsize 6 truein \epsfbox{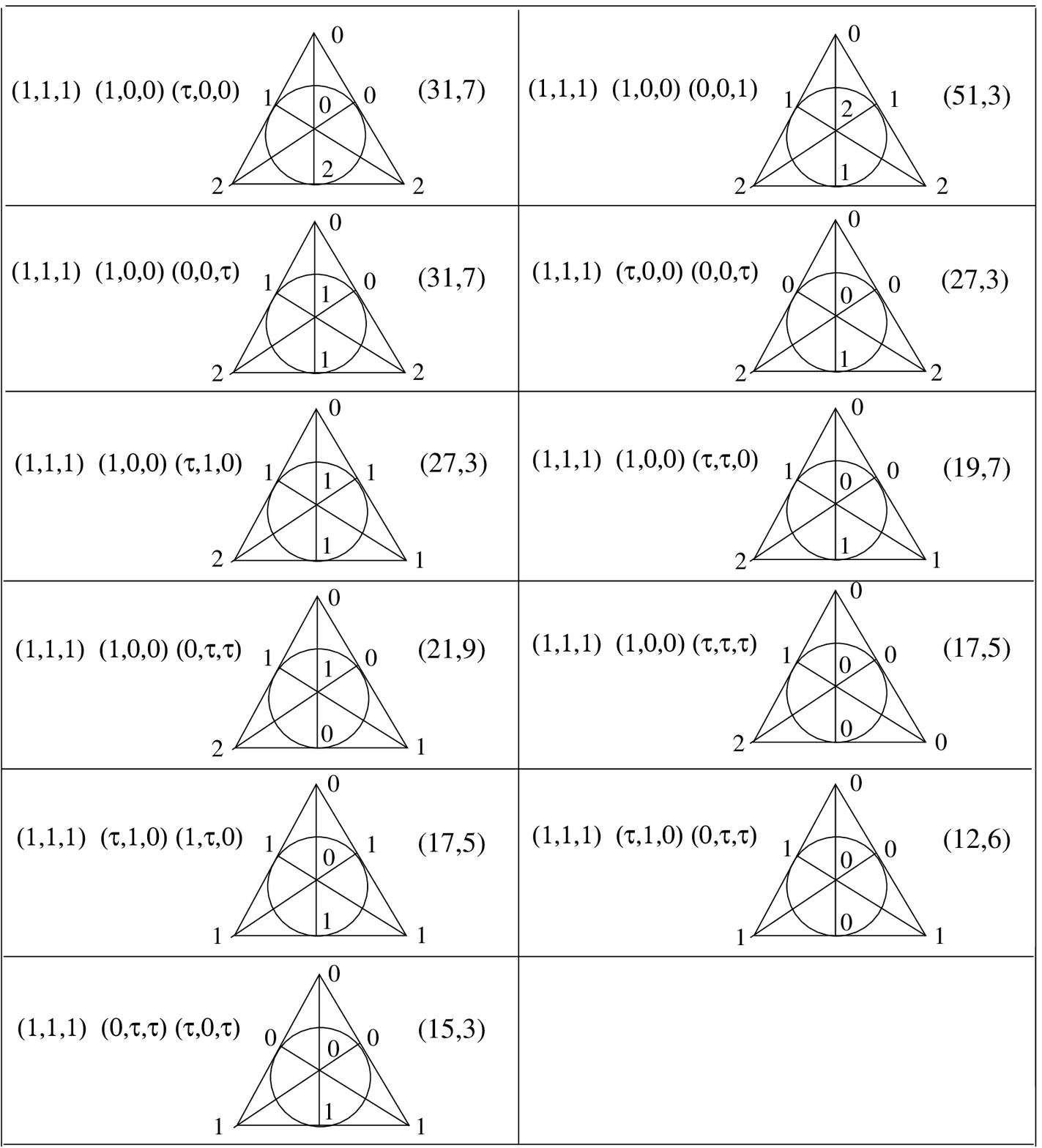}}
\caption{rank 3 classification}  
\label{figure3}
\end{figure}

The quotients of the $(27,3)$ model $X_2$ correspond
to subgroups ${\overline G}$ which contain the element
$(1,1,1)$. We can work out the complete list, which is
displayed in figure \ref{figure3}.

Clearly no two of these eleven subgroups can be equivalent,
as they all have distinct decorated diagrams. This classification can be
extended to cover the additional subgroups which do not contain $(1,1,1)$.
A cleaner way to see this will emerge shortly.

\section{Rules for orbifold cohomology
}\label{rulesfororco}
Having worked out some examples of groups $G$,
we need to compute the orbifold cohomology
of the quotient $T^6/G$. The general rule is:
\beq
H^*(T^6/G)=\oplus_{g\in G} H^*(T^g)^G
\eeq
Here $T^g$ is the fixed locus of $g\in G$ acting on $T^6$,
$H^*(T^g)$ is its cohomology, and the superscript $G$
denotes cohomology classes on $T^g$ which are invariant
under the action of $G$ on $T^g$.

We describe an element $g\in G$ by the data
$(\epsilon_1,\epsilon_2,\epsilon_3)(a_1,a_2,a_3)$.
Here each $\epsilon_i\in\pm1$, $\prod_{i=1}^3\epsilon_i=+1$,
and each $a_i$ is in $E_i[2]$. First let us describe the fixed
locus $T^g$:

$\bullet$ An element $g=(+,+,+)(a_1,a_2,a_3)\in {\overline G}\subset
G$ acts on $T^6$ by translations. Therefore $T^g=T^6$ if $g=0$,
and $T^g=\emptyset$ otherwise.

$\bullet$ An element $g=(-,-,+)(a_1,a_2,a_3)$ sends
$(x_1,x_2,x_3)\in T^6=E_1\times E_2\times E_3$ to
$(a_1-x_1,a_2-x_2,a_3+x_3).$ So the fixed points are\
of the form $(x_1={a_1\over 2},x_2={a_2\over 2},x_3$ arbitrary), 
and the fixed locus consists of $4\times 4=16$ copies of $T^2$,
if $a_3=0$; otherwise, $T^g=\emptyset$.

Next, we need the action of each element $h\in G$ on $T^g$,
in those cases where $T^g\neq\emptyset$.

$\bullet$ For $g=0$, we already saw that the $(Z_2)^2$ invariants
in $H^*(T^6)$ contribute a 3--dimensional space to
each of the Hodge groups $H^{11}$, $H^{21}$, $H^{12}$, $H^{22}$.
The action of any translation $g\in {\overline G}$ is trivial
on these spaces of invariants. We abbreviate this by saying that
the $g=0$ sector contributes $(3,3)$ to the orbifold cohomology. 

$\bullet$ We need the contribution to orbifold cohomology
of $g=(-,-,+)(a_1,a_2,0)$. Consider the subgroup
\beq
{\hat G}_3={\overline G}\cap(0,0,E_3[2])
\eeq
consisting of elements of ${\overline G}$ for which
$a_1=a_2=0$. Let $\rho_3$ denote the rank of ${\hat G_3}$;
it equals $0$, $1$ or $2$. Each of the $2^r$ translations
$h\in {\overline G}$ permutes the $16=2^4$ components
of $T^g$. This permutation is trivial if and only if
$h\in {\hat G}_3$. Therefore, the quotient of $T^g$
by the action of ${\overline G}$ consists of $2^{4-r+\rho}$
tori. 

We still have to impose invariance under $(Z_2)^2$.
Now the action of the twist $(-,-,+)$ on $T^g$ coincides
with the action of the translation $(a_1,a_2,0)$ for
which we have already accounted. So we are left solely
with invariance under the action of the $(-,+,-)$ twist.
This action sends a component labeled $({a_1\over 2},{a_2\over 2})$
to the component labeled $({a_1\over 2}+a_1,{a_2\over 2})$, {\it i.e.}
it shifts the labels by $(a-1,0)$. There are therefore two
possibilities: 

$\bullet$ If the projection of ${\overline G}$ to the 
first two factors $E_1[2]\times E_2[2]$ contains the
element $(a_1,0)$, then the action of $(-,+,-)$
relates two tori which had already been glued
previously by the action of the translation group
${\overline G}$. So it acts as the $\pm$ involution
on each of these $2^{4-r+\rho_3}$ components. The
result is $2^{4-r+\rho_3}$ copies of the quotient
$T^2/(\pm1)=P^1$. So the contribution to orbifold
cohomology is $(2^{4-r+\rho_3},0)$.

$\bullet$ Otherwise, the action of $(-,+,-)$ relates
pairs of tori which had previously been unrelated,
resulting in $2^{3-r+\rho_3}$ tori $T^2$.
The contribution to orbifold cohomology is therefore 
$(2^{3-r+\rho_3}, 2^{3-r+\rho_3})$.

To summarize, the contributions of the various sectors
to orbifold cohomology are: 

\beq
 \begin{tabular}{ccc}
$g=0$ ~~~~~~~~~~~~~~~~~~~~~~~~~~~~~~~~~~~~~~~~~~~~~~~& & $(3,3)$ ~~~\\
$g\in{\overline G},~g\neq0$ ~~~~~~~~~~~~~~~~~~~~~~~~~~~~~~~~~~~~~& &
$(0,0)$~~~\\
& & \\
$g=(-,-,+)(a_1,a_2,a_3)\Rightarrow$ ~~~~~~~~~~~~~~~~~~~~& & \\
  if $a_3\neq0$ ~~~~~~~~~~~~~~~~~~~~~~~~~~~~~~~~~& & (0,0) ~~~\\
  ~~~~~if $a_3=0$ and $(a_1,0,b)\in {\overline G}$ for some $b$ & &
~~~~~~$(2^{4-r+\rho_3},0)$ \\
  ~~~~if $a_3=0$ but $(a_1,0,b)\notin {\overline G}$ for any $b$ ~& &
~~~~~~~~~~~~~$(2^{3-r+\rho_3},2^{3-r+\rho_3})$ \\
\end{tabular}
\eeq
Finally, it is convenient to add the contributions of the four
elements $g$ corresponding to each ${\bar g}\in {\overline G}$
combined with the possible twists: 
\beq
 \begin{tabular}{ccc}
all $a_i\ne0$ ~~~~~~~~~~~~~~~~~~~~~~& &$(0,0)$~~~~~~~~~~~~~~~~~~~~~~~~~\\
$a_1\ne 0$, $a_2\ne 0$, $a_3=0$ ~~~& & \\
  ~~~~~~~~~~~~~~~~if $(a_1,0,b)\in {\overline G}$ for some $b$& &  
$(2^{4-r+\rho_3},0)$ ~~~~~~~~~~~~~~~~~\\
  otherwise ~~~~~& & $(2^{3-r+\rho_3},2^{3-r+\rho_3})$ ~~~~~~~~~\\
$a_1\neq0$, $a_2=a_3=0$ ~~~~~~~~& &$(2^{4-r}
(2^{\rho_2}+2^{\rho_3}),0)$~~~~~~~~\\
$a_1=a_2=a_3=0$ ~~~~~~~~~~~& &  ~~~~
$(3+2^{4-r}(2^{\rho_1}+2^{\rho_2}+2^{\rho_3}),3)$\\ 
\end{tabular}
\nonumber
\eeq
It is routine to apply these rules to each of the groups
$G$ encountered so far. The resulting Hodge numbers were tabulated 
in sections 4 and 5. 

\section{A reduction}\label{areduction}
In principle, we could list all subgroups of ranks 4, 5, 6
as we did for lower ranks, and we could compute their Hodge 
numbers according to the rules in section \ref{rulesfororco}.
However, there is a shortcut: If our translation 
subgroup subgroup ${\overline G}$ contains an element
$g=(a_1,a_2,a_3)$ where exactly one of the $a_i$'s (say $a_1$)
is non--zero then there is another subgroup ${\overline G}^\prime$
of lower rank such that the orbifolds $T^6/G$ and $T^6/G^\prime$,
where $G^\prime$ is the corresponding extension of ${\overline G}^\prime$,
are topologically equivalent, and in fact live in the same moduli space.
This reduces the calculation of the Hodge numbers for $G$ to those
for the smaller group $G^\prime$.

The reason for this is that the action of $(Z_2)^2$ commutes with the
translation $g$:

\begin{figure}[!h]
\centerline{\epsfxsize 2.0 truein \epsfbox{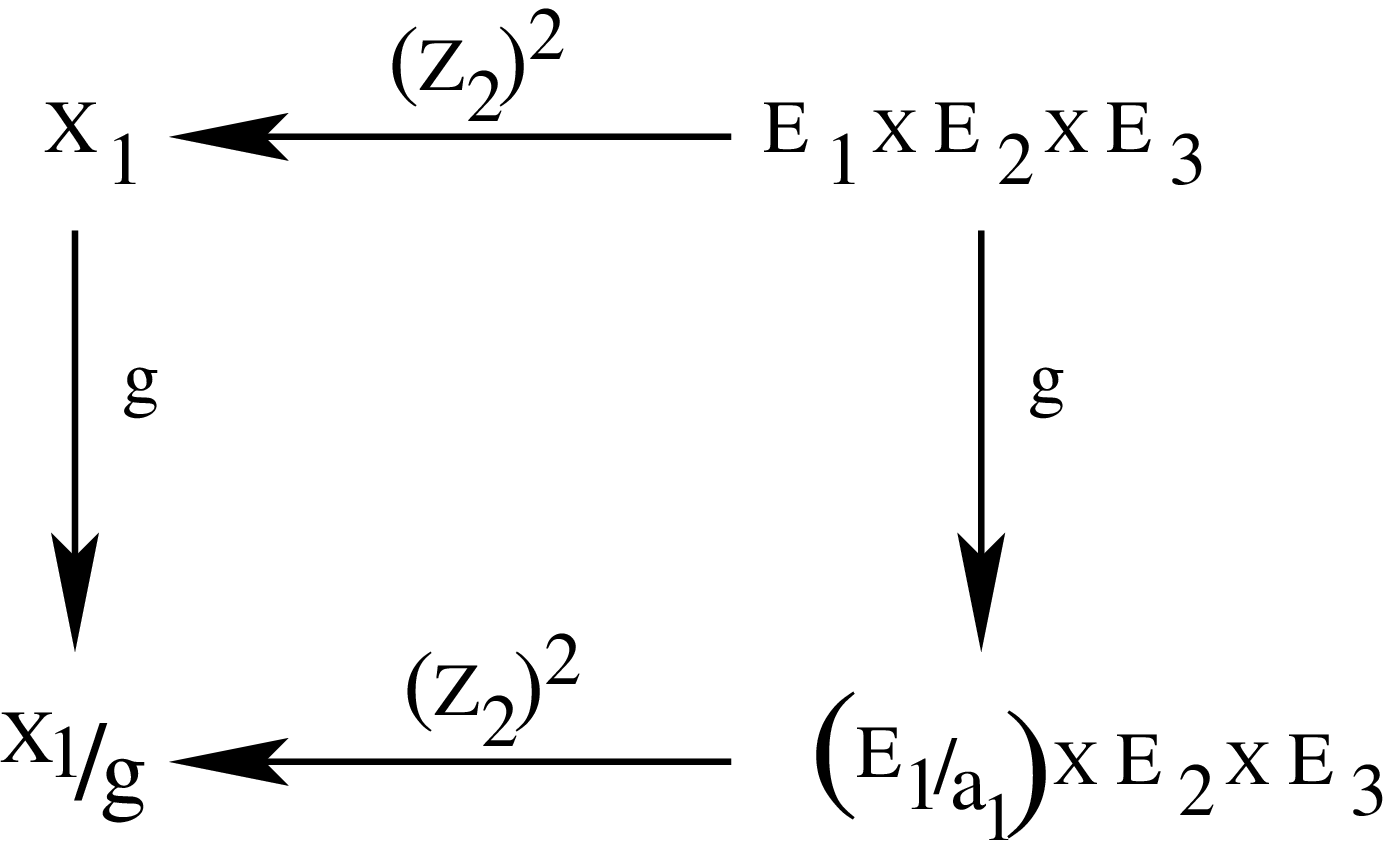}}
\label{aredu}
\end{figure}

Note that when more than one of the $a_i$ is non--zero,
we still have a commutative diagram, but the quotient
$(E_1\times E_2\times E_3)/g$ cannot be naturally identified
as a product of three $T^2$ factors. (In algebro geometric
language, the quotient is isogenous, but not isomorphic,
to the product of 3 elliptic curves.) 

We conclude that our $X=(E_1\times E_2\times E_3)/G=X_1/{\overline G}$
can also be described as $((E_1/a_1)\times E_2\times E_3)/{G^\prime}=
(X_1/g)/{\overline G}^\prime$, where ${\overline G}^\prime$
is the image of ${\overline G}$ which acts on $X_1/g$.
It is a subgroup of $(E_1/a_1)[2]\times E_2[2] \times E_3[2]$
of rank $r-1$. Since the Hodge numbers of $X=X_G$ depend
only on the group ${\overline G}$ (and not on the particular
elliptic curves used), we see that $H^*(X^G)=H^*(X^{G^\prime})$.
Inductively we can therefore assume that our group ${\overline G}$
contains no elements with a single non--zero entry.

\section{Rank $\geq~4$ 
}\label{beyondz2z2}


We are therefore led to study the groups of translations 
${\overline G}\approx Z_2^4$
contained in $Z_2^6=H^1(T^6,Z_2)$,
with the property that: $\forall~g\in~G$, $g\ne0$, 
$g=(a_1,a_2,a_3)$, $a_i\in E_i[2]$, at most one $a_i=0$.

We claim that
there exists such a group, and that it is in fact unique
up to natural equivalences.
We then take the (51,3) model modulo ${\overline G}$,
which is a $Z_2^3$ 
quotient of the (27,3) model. Our group
${\overline G}$ has 9 elements with
fixed points, each has a fixed point locus
$\approx Z_2\times $elliptic
({\it e.g.} $g=1/2(0,\tau,1)$ sends
$(x,y,z) \leftrightarrow (x,y+\tau/2,z+1/2)$,
so the fixed points are $(x,\tau/4,1/4)$,
which gives $4\times 4=16$ fixed tori mod 8 identifications). 
We still need to mod out by the $Z_2^3$ translations.
One of these acts trivially on its own fixed points.
The remaining $Z_2$'s act as follows: the first
interchanges the two tori, and the other acts as $z\rightarrow -z$,
so the fixed torus degenerates into ${\bf P}^1$. 
The resulting Hodge numbers are $(15,3)$.

We proceed to classify all such subgroups of $G$
which are generated by 4 vectors, one of which is (\ref{gammashift}).
In each $E_i$, the projection of $G$ must be all of $E_i[2]$.
Otherwise $G$ contains 8 elements projecting to 0 in $E_i$.
These 8 elements then form a hyperplane in $(Z_2)^4$,
given as perpendicular of some vector in $(Z_2)^4$, say $(a,b,c,d)$,
but then $(0,0,d,c)$ or $(b,a,0,0)$ is a non--zero 
vector in $Z_2^3$ with $0$ in another $E_j$. We conclude
that we can take the third and fourth basis vectors
of $G$ to project to 0 in $E_1$, and the second vector
projects to $\tau$
\beqn
G=1/2\{ 
        & (1,1,1); \nonumber\\
        & (\tau, x_3, a); \nonumber\\
        & (0,x_1,b);\nonumber\\
	& (0,x_2,c)\}
\label{genericform}
\eeqn
Now, $x_1$ and $x_2$ can be taken to be $\tau$ and $1$. Then by
subtracting multiples of the third and fourth vectors from the second, 
it follows
that we can take $x_3=0$. It remains to impose the conditions
on $a$, $b$ and $c$. These are
\beq
a\ne 0 ;  b\ne 0;  c\ne 0; c\ne b ;  c\ne a+1 ;  c\ne 1
\label{abcconditions}
\eeq
If these hold, no vector of $Z_2^4$ has $0's$ in $E$ and 
another factor, since the vectors with $0$ in $E$ are combinations 
of last two vectors and in columns 2,3 these each have 2 distinct
non zero entries. 
Last case to exclude is zeroes in column $2+3$. The vectors 
with zeroes in column 2 are $v_2$, $v_1+v_4$ and $v_1+v_2+v_4$. 
The third entries are then $a$, $1+c$, $1+a+c$, which impose the 
condition in eq. (\ref{abcconditions}).

We now proceed to compute the full set of solutions for $a$, $b$ and $c$.

\beq
 \begin{tabular}{|c|c|c|}
\hline
$a$ & $b$ & $c$ \\
\hline
1 & 1 & $\tau$ \\
1 & 1 & $\tau+1$ \\
1 & $\tau$ & $\tau+1$ \\
1 & $\tau+1$ & $\tau$ \\
$\tau$ & 1 & $\tau$ \\
$\tau$ & $\tau+1$ & $\tau$ \\
$\tau+1$ & 1 & $\tau+1$ \\
$\tau+1$ & $\tau$ & $\tau+1$ \\
\hline
\end{tabular}
\eeq
These solutions are invariant under two operations. One involves the
interchange of columns $E_1$ and
$E_2$ with $a\leftrightarrow b$ and $c\leftrightarrow c+1$. 
The second operation interchanges $\tau\leftrightarrow\tau+1$
and replaces $a\leftrightarrow a+c+1$, but only in column $E_1$. 
These two operations can be seen to mix all eight solutions and
therefore the solution is unique up to equivalences.

Since the rank 4 group is unique, it follows that every group
${\overline G}$ of higher rank can be reduced to rank 4
or less. 

\section{The complete list}\label{completelist}

Starting with the (51,3) model and analyzing the complete set of models
that are obtained by identifying fixed points on the three complex 
tori by shifts, we analyze the complete set of models that are obtained
from the $Z_2\times Z_2$ orbifold on a product of three complex tori.
The complete set of models is given in table (\ref{completeset}).

\beqn
& \begin{tabular}{|c|c|c|c|c|c|}
\hline
rank=1     & &  &  &  &  \\ 
\hline
(1,1,1) &                    &                   &                    &  
	$\Rightarrow$ (27,3) & 24\\
(1,1,0) &                    &                   &                    &  
	$\Rightarrow$ (31,7) & 24\\
\hline
rank=2     & &  &  &  &  \\ 
\hline
(1,1,1) & $(\tau,\tau,\tau)$ &                   &                    &  
	$\Rightarrow$ (15,3) & 12\\
(1,1,1) & $(\tau,\tau,0)$    &                   &                    &  
	$\Rightarrow$ (17,5) & 12\\
(1,1,1) & $(\tau,  1  , 0 )$ &                   &                    &  
	$\Rightarrow$ (19,7) & 12\\
(1,1,0) & $(0,1,1)$          &                   &                    &  
	$\Rightarrow$ (27,3) & 24\\
(1,1,0) & $(\tau,\tau,0)$    &                   &                    &  
	$\Rightarrow$ (21,9) & 12\\
\hline
rank=3     & &  &  &  &  \\ 
\hline
(1,1,1) & $(\tau,1,0)$       & $(1,\tau,0)$      &                    &
	$\Rightarrow$ (17,5) & 12\\
(1,1,1) & $(\tau,  1  , 0 )$ &  $(0,\tau,\tau)$  &                    &  
	$\Rightarrow$ (12,6) & 6\\
(1,1,1) & $(0, \tau, \tau)$  & $(\tau,0,\tau)$    &                    &  
	$\Rightarrow$ (15,3) & 12\\
\hline
rank=4     & &  &  &  &  \\ 
\hline
(1,1,1) & $(\tau,  0  , 1 )$ &  $(0,\tau,1)$     & $(0, 1 ,\tau)$ &  
	$\Rightarrow$ (15,3) & 12\\
\hline
\end{tabular}
\label{completeset}\\ \nonumber
\\ \nonumber
& {\rm Subgroups~of~} E[2]^3 {\rm ~free~of~single-entry~elements~~~}\nonumber
\\ \nonumber\
\eeqn

We note that the model with $|h_{11}-h_{21}|=3$,
that would correspond to the three generation case, does not
arise in this classification. We conclude that the 
$Z_2\times Z_2$ orbifold cannot produce three generations
solely with symmetric order 2 shift identifications on
the three internal complex tori. 

\section{The chirality condition}\label{chicon}

We now discuss a geometric picture of the chirality condition
(\ref{chiralitycondition}) that was discussed in section \ref{gs}.
We examine 
the fixed points of an element $(a,b,0)$, with $a$ and $b$ of order 2,
{\it i.e.} $(\epsilon + a/2, \delta+b/2,z)$. Under the action of the 
two twists the torus with parameters $(\epsilon, \delta)$ is shifted
under $(-,+,-)$ by $(a,0)$ and under $(+,-,-)$ by $(0,b)$.
The chirality question is whether one of these is a chirality projection
of a group element in $G$. Geometrically, we are trying to change the 
difference between the Hodge numbers, $h^{1,1}-h^{2,1}$. This can happen  
only when an involution acts on the $T^2$ above some fixed point as $-1$, 
so in the quotient this $T^2$ is replaced by a ${\bf P}^1$; this preserves 
$h^{1,1}$ but reduces $h^{2,1}$ by 1.

For an element of the form $(a,0,0)$, the fixed points in $T^6/(Z_2)^2$ are
$(a/2+\epsilon,z,\delta)$ and $(a/2+\epsilon,\delta,z)$.
In the resolution, these curves no longer intersect. The actions of the three 
non-trivial  elements are:

$(-,+,-)$: $(a,0,0)$ on either curve, with twist $z\rightarrow-z$ on
second curve. 

$(-,-,+)$: $(a,0,0)$ on either curve, with twist $z\rightarrow-z$ on
first curve. 

$(+,-,-)$: sends each curve to itself with a $z\rightarrow-z$.

We see that in this case the action is always chiral. 
We conclude that the quotient is non--chiral iff: 1) no element has 
two zeroes. 2) If an element has a zero, {\it e.g.}
$(a,b,0)$, then $(a,0,c)$ and $(0,b,c)$ are not in the group for any c.
This is therefore the translation of the 
chirality condition to the geometric language.

\section{Conclusion}\label{disco}

We have demonstrated in this paper that quotients of the 
$Z_2\times Z_2$ orbifold 
of a product of three $T^2$ tori by additional identifications
by shifts of order 2 on the three complex tori cannot 
reduce the net number of generations to three.
The motivation for this analysis stems from the
relation of these geometries to the free fermionic
heterotic string models and the manner in which three
generations are obtained in these string models.
Namely, there each one of the twisted sectors produces
a net number of one generation and it is therefore
of interest to explore the geometrical correspondence
of this picture. A complementary analysis, performed 
by using the free fermionic techniques \cite{fknr}, reaches
the same conclusion. Namely, the three generation cases
cannot be obtained by utilizing solely left--right symmetric shifts
on the complex tori, but necessarily involve an asymmetric
projection. This observation may have far reaching  implication
on the issue of moduli stabilization and vacuum selection. 
The reason is that the asymmetric operation cannot be performed
at an arbitrary point in the moduli space, but has to be performed
at special points. Extending the analysis of the $Z_2\times Z_2$
orbifold class models to the non--perturbative regime, along the 
lines of ref. \cite{mtheory}, will be facilitated by starting from
the $X_1$ manifold as the internal Calabi--Yau and solving the
anomaly constraints in the case of non--standard embedding
and in the presence of five branes. This will elucidate
how and whether the $Z_2\times Z_2$ reasoning for the origin of the
three generations is modified in the nonperturbative regime.
Similarly understanding
the implication of the asymmetric
operation in the context of strong--weak transformation
is of further interest. 
We should note that 
the type of geometries that correspond to
the left--right asymmetric action are not yet readily understood.

\section{Acknowledgments}

AF thanks the Mathematics department
at the University of Pennsylvania, and the Insititute for Advanced
Studies at Princeton, for hospitality. The work
of AF is supported in part by PPARC. RD is grateful for partial support from 
NSF grants DMS 0104354 and Focused Research Grant DMS 0139799.



\bigskip
\medskip

\bibliographystyle{unsrt}

\end{document}